\documentclass[amsfonts,twocolumn]{revtex4-1}
\usepackage{graphicx}
\usepackage{amsmath}
\usepackage{amsfonts}
\usepackage[dvips]{epsfig}
\usepackage[cm]{fullpage}
\usepackage{bm}
\usepackage[active]{srcltx}
\usepackage{epsf}
\begin{document}

\author{Shmuel Marcovitch and Benni Reznik}
\title{Structural unification of space and time correlations in quantum theory}
\affiliation{School of Physics and Astronomy,
Raymond and Beverly Sackler Faculty of Exact Sciences,
Tel-Aviv University, Tel-Aviv 69978, Israel.}
\begin{abstract}
We suggest a natural mapping between bipartite states and quantum evolutions of local states,
which is a Jamio{\l}kowski map.
It is shown that spatial correlations of weak measurements in bipartite systems
precisely coincide with temporal correlations of local systems.
This mapping has several practical and conceptual implications on the correspondence between Bell
and Leggett-Garg inequalities, the statistical properties of evolutions in large systems,
temporal decoherence and computational gain, in evaluation of spatial correlations of large systems.
\end{abstract}
\maketitle


Space and time are distinguished in the formalism of quantum theory.
A system that is separated in two parts of space is described by a positive semi-definite operator that lies in a tensor product of two Hilbert spaces.
The time evolution of a system is most generally described by a trace preserving map from one Hilbert space to another Hilbert space.
Mathematically, one can define a 
map 
between the space of
bipartite states and the space of time evolutions, which is
defined by the Hilbert-Schmidt scalar product.
However, one would not expect that this mapping would be physical,
that is that correlations of spatially separated observations would equal
the corresponding temporal correlations of measurements before and after the evolution.
In particular, while in the spatial case the measured operators commute,
in the temporal case two sequentially measured operators do not generally commute and
effect each other due to the uncertainty principle.

\begin{figure}[ht]
\center{
\includegraphics[width=3in]{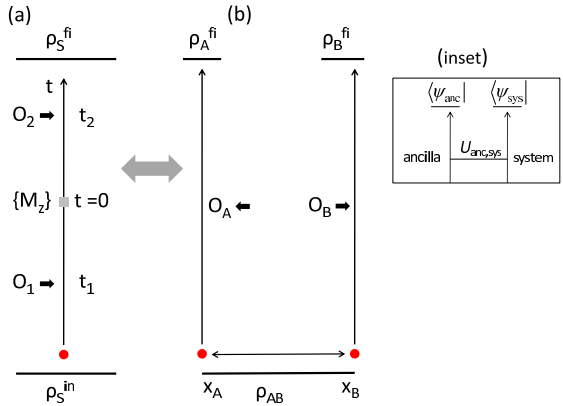}
\caption
{Mapping of bipartite states to time evolutions.
(a) A local state $\rho_S^{\text{in}}$ is weakly measured at $t_1<0$ and $t_2>0$ by $O_1$ and $O_2$ respectively,
where the system undergoes an instantaneous evolution given by Kraus operators $\{M_z\}$ at $t=0$. The system is post-selected to state $\rho_S^{\text{fi}}$.
(b) $A$ and $B$ share a bipartite state $\rho_{AB}$ and weakly measure it by $O_A$ and $O_B$ respectively.
Then the parties post-select their states to $\rho_A^{\text{fi}}$ and $\rho_B^{\text{fi}}$.
$\rho_{AB}$, $O_A(x_A)$, $O_B(x_B)$, $\rho_A^{\text{fi}}$ and $\rho_B^{\text{fi}}$ are mapped to
$\{M_z\}$, $O_1(t_1)$, $O_2(t_2)$, $\rho_S^{\text{in}}$ and $\rho_S^{\text{fi}}$, respectively.
\\Inset. Realization of post-selection to a mixed state by interaction with ancilla and post selecting both the system and the ancilla
to pure states.
}
\label{fig1}
}
\end{figure}
However, as is well known, there is a trade-off between
the accuracy of the measurement and the disturbance
caused to the system \cite{von}.
The limit in which individual measurements provide vanishing information gain
was first analyzed by Aharonov {\it et. al.} \cite{weak} 
and was termed {\it weak measurements}.
Since weak measurements only slightly disturb the systems,
they provide a non-destructive and operational method for comparing spatial and
temporal correlations in quantum mechanics.

In this letter we construct a Jamio{\l}kowski map \cite{jam} between the space of bipartite systems
$\rho_{AB}\in H_A\otimes H_B$
and the set of time evolutions transforming systems from $H_A$ to $H_B$. 
In this mapping spatial correlations of weak measurements in bipartite systems
precisely coincide with temporal correlations of weak measurements, before and after the evolution, in local systems (Theorem 1).
The entanglement between $A$ and $B$ is mapped to a correlation between the past and the future,
which characterize the evolutions of systems and their quantum mechanical nature.
We show that maximally entangled states are mapped to unitary evolutions.
Non-maximally entangled states correspond to evolutions under the influence of selective measurement.
In particular, non-entangled pure product states correspond to selective projector measurements.
Finally, mixed bipartite systems are mapped to mixtures of the corresponding evolutions.
We shall also discuss briefly several practical and conceptual applications of the suggested mapping.


To set the ground for the mapping let us first discuss generalized time evolutions.
The evolution of a system $\rho_S$, subject to interaction with a larger system,
is most generally described as a completely positive map given by
Kraus operators $\{M_z\}$ \cite{kraus}:
$\rho_S\rightarrow\sum_{z}M_z\rho_S M_z^{\dagger}$,
where
$\sum_z M_z^{\dagger} M_z=I$.
Beyond the trivial unitary operations Kraus operators describe evolutions
due to the interaction with an environment.

In order to describe the effect of selective measurements, we remove the constraint $\sum_z M_z^{\dagger} M_z=I$ 
and normalize the Kraus operators
to preserve the trace:
\begin{equation}
\label{normalization}
M_z\rightarrow M'_z=\frac{\sqrt{p_z}M_z}{\sqrt{\text{Tr}[\sum_{z'} p_{z'} M_{z'}^{\dagger} M_{z'} \rho_S]}},
\end{equation}
where $p_z\geq0$ is the probability for post-selecting the state dictated by $M_z$. 
A single Kraus operator corresponds to selecting the state $M'^\dagger \rho_S M'$ ($p=1$).

As an illustration, consider the two-dimensional case in which the system $\rho_S$
is measured in the computational basis non-selectively: $M_0=|0\rangle\langle0|$ and $M_1=|1\rangle\langle1|$.
$\rho_S$ then evolves to the diagonal form
$\rho\rightarrow \langle0|\rho_S|0\rangle M_0+\langle1|\rho_S|1\rangle M_1$.
Alternatively, if $\rho_S$ is subject to selective measurements $M_0$ with probability $p_0$
and $M_1$ with probability $p_1$,
by normalizing $M_0$ and $M_1$ according to Eq. (\ref{normalization}) one obtains a more general evolution
$\rho_S\rightarrow q_0 M_0+q_1 M_1$, where
$q_z=p_z\langle z|\rho_S|z\rangle/(\sum_{z'} p_{z'}\langle {z'}|\rho_S|{z'}\rangle)$,
which coincides with the non-selective case in case all $p_z$ are equal.


In the following we present the main results.
\\{\bf Temporal weak correlations.}
An initially prepared system $\rho_S^{\text{in}}$ with dimension $d_A$ is subject to an evolution described by Kraus operators
(as normalized in Eq. \ref{normalization}),
which for the sake of simplicity we take as instantaneous at time $t=0$.
We assume that the system is measured weakly (and instantaneously) before $t=0$ and after $t=0$
by operators $O_1$ and $O_2$ with two pointer readings $q_1(t_1)$ and $q_2(t_1)$ respectively,
as illustrated in fig. 1(a).
\\{\bf Lemma 1.} {\it
The correlation of the instruments' pointers $q_1(t_1)$ and $q_2(t_1)$ is given by:}
\begin{equation}
\label{temporal}
E(q_1 q_2 )=\frac{1}{2}\text{Tr}\left[O_2\,\sum_z M'_z\{O_1,\rho_S^{\text{in}}\}M_z^{'\dagger}\right],
\end{equation}
which includes both selective and non-selective measurements.
See related results in the context of unitary evolutions 
for correlations of two-level system with continuous weak measurements \cite{korotkov}, in the context of post-selection \cite{steinberg,jozsa,stein2} and of two sequential measurements \cite{johansen,ours}.
\\{\bf Lemma 2.} {\it By also post-selecting to state $\rho_S^{\text{fi}}$ with dimension $d_B$ }
\begin{equation}
\label{post}
E(q_1 q_2) = \frac{1}{4}\text{Tr}\left[\rho_S^{\text{fi}}\left\{O_2,\sum_z M'_z\{O_1,\rho_S^{\text{in}}\}M^{'\dagger}_z\right\}\right],
\end{equation}
where
\begin{equation}
\label{C2}
M'_z=\frac{\sqrt{p_z} M_z}{\sqrt{\text{Tr}\left[\rho_S^{\text{fi}}\sum_{z'} p_{z'} M_{z'} \rho_S^{\text{in}} M^{\dagger}_{z'}\right]}}.
\end{equation}
Post-selection to a mixed state using ancillas is described in the inset of fig. 1.
Note that no post-selection is equivalent to post-selecting the maximally distributed mixed state $I/d_B$.
\\{\bf Spatial weak correlations.}
Next, let us assume an initially prepared bipartite system $\rho_{AB}$ is measured weakly by
parties $A$ and $B$ with operators $O_A$ and $O_B$ respectively, as illustrated in fig. 1(b).
In addition, $A$ may post-select her state to $\rho_A^{\text{fi}}$ and $B$ to $\rho_B^{\text{fi}}$.
An immediate consequence of Eqs. (\ref{temporal},\ref{post}) is
\\{\bf Corollary 1}. {\it The correlation of spacelike related pointers  $q_A(x_A)$ and $q_B(x_B)$ equals:}
\begin{equation}
\label{spatial}
E(q_A q_B) = \frac{\text{Tr}\left[\rho_A^{\text{fi}} \otimes \rho_B^{\text{fi}}\left\{I_A\otimes O_B,\{O_A\otimes I_B,\rho_{AB}\}\right\}\right]}
{4\text{Tr}\left[ (\rho_A^{\text{fi}}\otimes \rho_B^{\text{fi}} )\rho_{AB}\right]}.
\end{equation}
Note again that for each party no post-selection is equivalent to post-selecting the maximally distributed mixed state.

Finally, let us present the mapping between time evolutions and bipartite states.
A pure bipartite state is mapped to a single
Kraus operator by having
\begin{equation}
\label{M}
|\psi_z\rangle=\sum_{ij}\alpha^z_{ij}|i\rangle\otimes|j\rangle\quad\Leftrightarrow\quad M_z=\sum_{ij}\alpha^{z*}_{ji}|i\rangle\langle j|.
\end{equation}
where $d_A$ is the dimension of $H_A$, $1\leq i\leq d_A$, and $d_B$ of $H_B$, $1\leq j\leq d_B$.
The map extends to mixed states/evolutions by convex combinations:
\begin{equation}
\label{M2}
\rho_{AB}=\sum_z p_z |\psi_z\rangle\langle\psi_z|\quad\Leftrightarrow\quad \rho_S\rightarrow \sum_z p_z M_z^{'\dagger} \rho_S M'_z.
\end{equation}
{\bf Theorem 1}. {\it Given the mapping defined in Eqs. \ref{M}, \ref{M2} and the following correspondence of operators and boundary states:
$O_1=O_A$, $O_2=O_B^*$, $\rho_S^{\text{in}} = \rho_A^{\text{fi}}$, and $\rho_S^{\text{fi}}=\rho_B^{fi*}$,
}
\begin{equation}
\label{main}
E\big(q_1(t_1) q_2(t_2)\big)=E\big(q_A(x_A) q_B(x_B)\big).
\end{equation}
This mapping is illustrated in figure 1.
It is symmetric to the exchange of $A$ and $B$, given
that we exchange the dimensions of the boundary conditions of $\rho_S$ and take $M_z^{\text{t}}$ instead of $M_z$.
Note that the usual spatial setting in which no post-selection is assumed, corresponds to having
a maximally distributed mixed state in the temporal setting $\rho_S^{\text{in}}=I/d_A$ with no post-selection.
\\ {\bf Corollary 2}. {\it The expectation values of the single measurements equal as well:}
$E(q_1)=E(q_A)$, $E(q_2)=E(q_B)$.

It is illuminating to analyze points in which our mapping does not work.
The notion of multipartite entanglement is well established by now.
One may then expect that tripartite correlations would also be mapped to temporal ones.
This is not the case, however, even in the simplest case where the state $\rho_S^{\text{in}}$ evolves trivially ($M=I$) and not post-selected.
It can be shown that the correlation of three sequential weak measurements $O_1$, $O_2$ and $O_3$ performed at times $t_1<t_2<t_3$ is given by
\begin{equation}
\label{tripartite}
E(q_1q_2q_3)=\frac{1}{4}\text{Tr}\left[O_3,\{O_2,\{O_1,\rho_S^{\text{in}}\}\}\right].
\end{equation}
In contrast to measurements at two times, Eq. (\ref{tripartite}) implies that the correlation of three {\it depends on their order},
in sharp contradiction with the multipartite spatial scenario.
This reflects that one dimensional time can only be bisected once to unordered parts,
whereas multidimensional space may be sectioned into many parts with no internal order.
Another feature of multipartite states which is not satisfied in the temporal setting is
the monogamy of entanglement \cite{wootters2}.
If two qubits $A$ and $B$ are maximally entangled, they cannot be correlated at all with a third qubit $C$.
In the temporal case, however, we choose again $M=I$.
Then any pair of instances among $t_1$, $t_2$, $t_3$ {\it etc.} is maximally correlated.

We would like to remark that a notion of {\it entanglement in time} was introduced in a different context by Brukner {\it et. al.} \cite{vedral},
who analyze correlations of successive $\pm1$ strong measurements.
These temporal correlations violate Leggett-Garg inequalities \cite{leggett},
the Bell inequalities \cite{bell} in time.
Brukner {\it et. al.} also show that there are no genuine multi-time correlations 
and that the monogamy of spatial correlations is violated in the temporal setting.
However, there are crucial differences between temporal correlations of strong and
weak measurements as
correlations of successive strong measurements do not depend on the state and are a particular feature of $\pm1$ observables.
The suggested mapping does not apply for strong measurements.

We note that different physical interpretations of Jamio{\l}kowski isomorphism were given:
a purification protocol of local unitary operations \cite{cirac}, 
a test of non-zero channel capacity \cite{Nowakowski}, and a manifestation of
"superposition of unitary operations" \cite{benni}.

We proceed by proving our results.
\\{\bf Proof of Lemma 1. }
Observables $O_1,\ O_2$ are measured sequentially on system $\rho_S^{\text{in}}$
at times $t_1, t_2$ where $t_1<0<t_2$.
In addition, $\rho_S^{\text{in}}$ evolves at $t=0$
with Kraus operators $\{M_z\}$.
The von-Neumann interaction measurement corresponding to $O_1$ and $O_2$ is
$H_{int}=\delta(t-t_1)p_1 O_1+\delta(t-t_2)p_2 O_2$,
where $[q_i,p_i]=i (\hbar=1)$.
We assume identical initial Gaussian wavepackets $\phi(q_1)$ and $\phi(q_2)$ for the pointers:
\begin{equation}
\rho_i=\phi(q_i)\phi(q'_i)=\int dq_i dq'_i \sqrt{\frac{\epsilon}{2\pi}}e^{-\epsilon (q_i^2+q_i^{'2})/4}, (i=1,2).
\end{equation}

The initial state of the system and the apparatuses $\rho_S^{\text{in}}\otimes\rho_1\otimes\rho_2$,
evolves to
$U_2\left[\sum_z M'_z(U_1 \,\rho_S^{\text{in}} \otimes \rho_1\otimes\rho_2\, U_1^{\dagger})M_z^{'\dagger}\right]U_2^{\dagger}$,
where
$U_i=e^{-i p_i O_i}$ (We assume $H=0$ at times different from $t_1$, $0$ and $t_2$).
Each operation of $p$ yields an order of $\epsilon$, where in the limit of weak measurements $\epsilon\to0$.
By expanding $U_i$ to second order ($i=1,2$)
$U_i=1-i p_iO_i-\frac{1}{2}p_i^2O_i^2+o(\epsilon^3)$,
one can compute the composite state of the system and pointers:
\begin{widetext}
\begin{equation}
\label{expand}
\begin{split}
& \rho_S^{\text{in}}\otimes\rho_1\otimes\rho_2\ \rightarrow \ \sum_z M'_z\rho_S^{\text{in}} \rho_1\rho_2M_z^{'\dagger}
 - \sum_z M'_z\left\{O_1,\rho_S^{\text{in}}\right\} M_z^{'\dagger}\phi'(q_1)\phi(q_1')\rho_2
 +\frac{1}{2}\sum_z M'_z\left\{O_1^2,\rho_S^{\text{in}}\right\} M_z^{'\dagger}\phi''(q_1)\phi(q_1')\rho_2
\\& \!-\! \left\{\!O_2,\!\sum_z M'_z \rho_S^{\text{in}} M_z^{'\dagger}\!\right\}\!\phi'(q_2\!)\phi(q_2')\rho_1
\!+\!\frac{1}{2}\!\left\{\!O_2^2,\!\sum_z M'_z \rho_S^{\text{in}}M_z^{'\dagger}\!\right\}\!\phi''\!(q_2\!)\phi(q_2')\rho_1
\!+\!\left\{\!O_2,\!\sum_z M'_z\!\left\{\!O_1,\rho_S^{\text{in}}\!\right\}\! M_z^{'\dagger}\!\right\}\!\phi'(q_1\!)\phi(q_1')\phi'(q_2\!)\phi(q_2').
\end{split}
\end{equation}
\end{widetext}
To compute $E(q_1 q_2)$ we first notice that since
$\int q\phi^2(q)dq=0$ and $\int\phi(q)\phi'(q)dq=0$,
all terms in Eq. (\ref{expand}) except the last one do not contribute.
In addition, by tracing out the system $\rho_S^{\text{in}}$ and using $\int q \phi(q)\phi'(q)dq=-1/2$,
we conclude that
$E(q_1 q_2) = \frac{1}{4}\text{Tr}\left[\left\{O_2,\sum_z M'_z\left\{O_1,\rho_S^{\text{in}}\right\} M_z^{'\dagger}\right\}\right]$,
which coincides with Eq. (\ref{temporal}).
\\{\bf Proof of Lemma 2. }
Preparation of a mixed state is realized by projecting a system to a pure state $|\psi_S^{\text{in}}\rangle$
which then interacts with an ancilla in a known state. 
Correspondingly, post selection to a mixed state $\rho_S^{\text{fi}}$ is realized in the same way but
with the reversed time axis:
$\rho_S^{\text{fi}}=U_{int}^{\dagger}|0_{anc}\rangle\langle0_{anc}|\otimes|\psi_S^{\text{fi}}\rangle\langle\psi_S^{\text{fi}}|U_{int}$
(as illustrated in the inset of figure 1).
The proof of Lemma 2 follows the same steps as that of Lemma 1 where instead of tracing out the system,
one projects the system to the final state and renormlizes the remaining state.
In case $M=I$ the normalization yields a factor of
$1/\text{Tr}[\langle 0_{anc}|\langle \psi_S^{\text{fi}}|U_{int}\rho_S^{\text{in}}\otimes I_{anc} U_{int}^{\dagger}|\psi_S^{\text{fi}}\rangle|0_{anc}\rangle]=1/\text{Tr}[\rho_S^{\text{in}}\rho_S^{\text{fi}}]$.
The generalization to arbitrary evolution is straightforward.
Note that Wizeman \cite{wiseman} analyzed a similar case 
for a single weak measurement.
%
\\{\bf Proof of Theorem 1. } 
Let us first show the correspondence for a pure bipartite state $|\psi\rangle$, which is mapped to a single Kraus operator $M'$ (with $p=1$).
We show the equality of the temporal and spatial denominators $D_T$, $D_S$ and nominators $N_T$ and $N_S$ of Eq. (\ref{post}) and (\ref{spatial}) respectively.
From Eq. (\ref{C2},\ref{M}) up to a factor of $4$:
$D_T=\text{Tr}\left[\rho_S^{\text{fi}} M \rho_S^{\text{in}} M^{\dagger}\right]
=\sum_{i,j,k,l} \alpha_{ij}\alpha^*_{kl}\rho_{S_{ki}}^{\text{in}}\rho_{S_{ij}}^{\text{fi}}$,
$D_S=\text{Tr}\left[(\rho_A^{\text{fi}}\otimes \rho_B^{\text{fi}})|\psi\rangle\langle \psi |\right]
=\sum_{i,j,k,l} \alpha_{ij}\alpha^*_{kl}\rho_{A_{ki}}^{\text{fi}}\rho_{B_{lj}}^{\text{fi}}$,
$N_T\!=\!\text{Tr}\!\left[\!\rho_S^{\text{fi}}\!\{O_2,M\{O_1,\rho_S^{\text{in}}\}\! M^\dagger \}\!\right]\!
=\!\sum_{i,j,k,l}\!\alpha_{ij}\alpha^*_{kl}\{\rho_S^{\text{in}},O_1\}_{ki}\{\rho_S^{\text{fi}},O_2\}_{jl}$,
and $N_S\!=\!\text{Tr}\!\left[\!(\rho_A^{\text{fi}}\!\otimes \!\rho_B^{\text{fi}})\!\{I_A\otimes O_B,\{O_A\otimes I_B,|\psi\rangle\langle\psi|\}\!\}\!\right]\!
=\!\sum_{i,j,k,l}\alpha_{ij}\alpha^*_{kl}\{\rho_A^{\text{fi}},O_A\}_{ki}\!\{\rho_B^{\text{fi}},O_B\}_{jl}$,
where we use the notation $A_{ki}=\langle k|A|i\rangle$ for matrix elements.
In correspondence with the mapping defined in Theorem 1, $D_T=D_S$ and $N_T=N_S$.
Note that by proving $D_S=D_T$ we have explicitly confirmed that the mapping corresponds to Jamio{\l}kowski isomorphism \cite{jam}.
To extend to a set of Kraus operators $M'_z$, note that $D_T$, $D_S$, $N_T$, $N_S$ become now a convex combinations of $p_z$,
which respects their equality. This concludes the proof of Theorem 1. $\square$

{\bf Implications.}
The suggested mapping has several important implications which we briefly discuss:
\\1. {\it The correspondence between Bell and Leggett-Garg inequalities}. 
Interestingly, Leggett and Garg \cite{leggett} have suggested temporal inequalities 
with the same bounds as the corresponding spatial Bell inequalities \cite{bell}.
For example, CHSH inequality \cite{chsh} and the corresponding temporal inequality (Eq. 2b in \cite{leggett}) are bounded by $2\sqrt{2}$ \cite{tsirelson}.
In a previous paper \cite{ours} we have shown that Bell's inequalities can be maximally violated using weak measurements even if
all observables are measured for each member of the ensemble.
A similar result for Leggett-Garg inequalities was given in \cite{mizel,jordan1}.
By the mapping above the correspondence between the two type of inequalities becomes clear.
Leggett-Garg inequalities are distinguished from the Bell inequalities as their maximal violation depends only on
the measured observables and not on the state of the system.
By the mapping above we see that this is a consequence of unitary evolutions which correspond to maximally entangled bipartite states.
By having non unitary evolutions Leggett-Garg inequalities are less violated.

Since our mapping is exact all the results concerning bipartite Bell inequalities are valid in the corresponding temporal inequalities.
For example, the non-separable Werner states which do not violate CHSH inequality \cite{werner},
correspond to the same mixtures of unitary evolutions which do not violate Leggett-Garg inequality.
Another example is the anomaly of nonlocality in bipartite systems with dimension greater than two \cite{anomaly},
with a Bell inequality \cite{g_bell2,g_bell2add}
that is not maximally violated by the maximally entangled state.
One can explicitly show that the same anomaly appears in the temporal setting,
where maximal violation is obtained with the corresponding non-unitary evolution.
\\2. {\it Statistical characteristics of large systems}. In the work by Hayden {\it et. al.} \cite{popescu_thermo}
correlation properties of random high-dimensional
bipartite pure systems were examined.
They showed that there exist large subspaces in which almost all pure states
are close to maximally entangled.
Their result is based on the uniqueness of the Haar measure in pure states.
Any pure state can be generated by applying a unitary matrix on a fiducial state,
where the space of unitary matrices is comprised of the rotationally invariant Haar measure.
Through our mapping the space of bipartite unitary evolutions maps also to "pure evolutions" on local systems where a pure evolution
corresponds to a single Kraus operator. 
Therefore, there exist large subspaces in which all pure evolutions are close to unitary ones.
This implies that as the system becomes sufficiently large, its evolution
is most likely to be unitary.
\\3. {\it Models of decoherence}. The usual framework of decoherence \cite{lindblad} deals with the transition of a state to a diagonal form.
By the suggested mapping
one can distinguish decoherence of states from {\it decohering dynamics}.
Decohering dynamics can be observed by detecting the temporal decay of correlations in case of non exact unitary evolution,
even on the maximally distributed mixed state.
\\4. {\it Computational gain}. In numerical computations of two point correlation function of bipartite states, needed
for instance in evaluating Bell inequality bounds, one can utilize the corresponding
Leggett-Garg inequalities. 
For example, given $d_A=d_B=N$, instead of manipulating $N^2\times N^2$ matrices, one can use only $N\times N$ matrices.
\\{\bf Conclusions.} Our mapping provides a new perspective on time evolution in quantum-mechanics.
By observing correlations of weak measurements the entanglement of bipartite states finds exact correspondence with temporal quantum mechanical correlations.
Surprisingly, by having an exact mapping between spatial and temporal correlations,
non-relativistic quantum-mechanics manifests a structural unification of time and space.

\acknowledgments
We are deeply grateful to Y. Aharonov whose insights initiated this work. We also thank A. Botero and P. Skrzypczyk.
This work has been supported by the Israel Science Foundation grant number 920/09, the German-Israeli foundation, and the European Commission (PICC).

\end{document}